\documentclass{article}
\usepackage[a4paper, margin=1.2in]{geometry}
\usepackage{graphicx}
\usepackage{amsmath}
\usepackage{array}
\usepackage{color}
\usepackage{rotating}
\usepackage{hyperref}
\usepackage{subcaption}
\usepackage[square,sort,comma,numbers]{natbib}

\begin{document}

\title{\Large Characterizing Transactional Databases for Frequent Itemset Mining}
\author{Christian Lezcano\thanks{Computer Science Department, Universitat Polit\`{e}cnica de Catalunya, Spain, clezcano@cs.upc.edu} 
\and Marta Arias\thanks{Computer Science Department, Universitat Polit\`{e}cnica de Catalunya, Spain, marias@cs.upc.edu}}
\date{}

\maketitle

\begin{abstract} \small\baselineskip=9pt This paper presents a study of the characteristics of transactional databases used in frequent itemset mining.
Such characterizations have typically been used to benchmark and understand the data mining algorithms working on these databases.
The aim of our study is to give a picture of how diverse and representative these benchmarking databases are, both in general but also in the context of particular empirical studies found in the literature.
Our proposed list of metrics contains many of the existing metrics found in the literature, as well as new ones. Our study shows that our list of metrics is able to capture much of the datasets' inner complexity and thus provides a good basis for the characterization of transactional datasets. Finally, we provide a set of representative datasets based on our characterization that may be used as a benchmark safely.\end{abstract}

\section{Introduction} \label{intro}

Since the introduction of Frequent Itemset Mining (\textit{FIM}) and its early algorithms, a huge number of algorithms have been proposed \cite{apriori,Han2004,Zaki:2000:SAA:627328.628066,989550,Uno2004LCMV2} (see \citet{DBLP:journals/widm/Fournier-VigerL17} for a more detailed review of the latest \textit{FIM} approaches); in fact, itemset mining and the closely related association rule mining have been arguably the hottest topic within the field of data mining for years.
With the appearance of competing itemset mining algorithms comes the need to understand their strengths and weaknesses. Natural questions arise: what algorithm is the fastest for one particular dataset\footnote{We use dataset and database interchangeably.}? What is the best algorithm? Or more realistically: what algorithms work best for what types of datasets?

In an attempt to answer these questions, authors set up and run empirical studies (what we call \emph{algorithm benchmarking} here).
In data mining algorithm benchmarking, one uses a set of datasets as the basis for comparison (a \emph{benchmark}), and applies all competing algorithm candidates to the benchmark in order to establish what algorithm is the fastest, uses less memory or gives more accurate results. In order to make the comparison fair, one should use as many datasets as possible, and these should be as diverse as possible. In other words, the benchmark should be a representative sample of what is to be expected when applying the algorithms in real-world scenarios.

The question of whether a benchmark (that is, a set of databases used for comparing algorithms) is representative or not is a difficult one to answer. 
Our proposed approach in this work is to characterize each dataset with the computation of several metrics, each capturing a different aspect of the datasets' complexity and structure. Trivial metrics are for example the number of transactions in the database, and more complex ones are the ones establishing, for example, their \emph{density} \cite{Gouda:2001:EMM:645496.658047}.
Once one has a description of each dataset by means of a vector of metrics, one can see how the different metrics' values are distributed. If they cover a wide range collectively, then the benchmark is a good candidate for being representative. If, on the other hand, values are all clumped together, then the benchmark is probably not a good one to use.

Another use for the characterization of databases is to establish connections between database characteristics and the performance of particular data mining algorithms, so that one could for example make claims such as ``algorithm A works well for databases that have many transactions and are dense, but algorithm B works better if the database is sparse and small.''

In fact, there are several works that do precisely this, namely, establishing connections between dataset characteristics and algorithm performance \cite{Zheng:2001:RWP:502512.502572,DBLP2003}.
We detail these works in the next section.

An important tool that we exploit in this paper is the \textit{FIMI} public repository, where \textit{FIM} algorithms and benchmark datasets were made publicly accessible to the \textit{FIM} community. Since its introduction by \citet{DBLP2003}, this repository serves as a standard set of tools to benchmarking algorithmic approaches. It has been very successful in the sense that most benchmarking studies use datasets from this repository.
Another public repository of transactional datasets is the \textit{SPMF} open-source data mining mining library. This repository is constantly updated and provides a greater number of \textit{FIM} implementations and datasets \cite{10.1007/978-3-319-46131-1_8}.
We believe that is paramount to characterize the transactional databases present in both repositories. For that reason, we use metrics found in the literature and others of our own in order to understand their nature and evaluate their representativeness.

Another area where characterizing databases may be of use is in synthetic database generation. When attempting to generate synthetic data, one needs some control over the generated data. This control may come in the shape of characterizing the data generated, for example, by means of metrics similar to the ones we propose here.

Finally, we believe that taking into account the characterization of databases when making claims about an algorithm's performance over competitors is important in order to give enough support to such claims. We recognize the fact that this may be out of the scope of some work where the focus is on algorithm development. To make the life of such developers easier, we propose a benchmark that covers the full spectrum of values found in our study as a ``minimum representative benchmark'' (MRB), so that authors that use our proposed MRB have some guarantees that the databases used are representative. Naturally, if in the future new metrics emerge, the list may have to be revised, so we propose this as an evolving MRB. 
To summarize, the contributions in this paper are:

\begin{enumerate}
\item We provide a comprehensive list of metrics used for database characterization
\item We evaluate the metrics mentioned in the item above over publicly available and commonly used databases used for frequent itemset mining algorithms
\item We define and study benchmark representativeness of existing works
\item We propose a  minimum representative benchmark, i.e., a benchmark whose representativeness is guaranteed according to the metrics included here
\end{enumerate}

The organization of this paper is as follows. Section 2 describes the related work which is divided into three parts. Section 3 presents the definition of the metrics used in the characterization of transactional datasets. Then, in Section 4 the experiments and evaluations carried out in this work are presented. Finally, Section 5 explains the conclusion and future work.

\section{Related work}
\label{related}

\paragraph{On the connections between database characteristics and algorithm's performance.}
To the best of our knowledge, the work by \citet{Zheng:2001:RWP:502512.502572} was the first one to note the relevance of data characteristics over an algorithm's performance. The authors show that algorithms that appear better suited for some specific dataset properties do not respond in the same way for other classes of datasets. Specifically, they report that algorithms measured over synthetic datasets behave very differently when run over real datasets. The same authors explain that the reason for this outcome might be that the algorithms were fine-tuned for the synthetic dataset characteristics used in their experiment which caused to perform poorly on the real-life datasets. Since then there has been awareness of the importance of benchmarking methods over databases with different set of dataset characteristics. 
\citet{DBLP2003} address the problem claimed by \citet{Zheng:2001:RWP:502512.502572} and introduce the $\mathit{FIMI}$ repository with much success.

\paragraph{On algorithm benchmarking.}
The work of \citet{zimmermann2015data} points out several existing issues in the evaluation of frequent itemset mining algorithms. Here, the author denounces the lack of the necessary diversity of characteristics to fully understand the algorithms' strengths and limitations. The author notes that it is not the number of datasets utilized in measuring the quality of an algorithm that matters, but it is evaluating an algorithm with datasets with a variety of characteristics representing real world scenarios. On the other hand, it is also noted by the aforementioned author that simply adding more datasets to a repository does not entail that more different characteristics are being added. Briefly, the problem above motivates the use of a collection of real datasets or artificial dataset generators that emulate the whole spectrum of genuine characteristics found in real-life databases.
In the same way, it is explained that every dataset utilized in benchmarking should be comprised of characteristics which emerge from a real-life process. This means that merely generating new datasets under randomly selected characteristics can not be considered appropriate since these datasets do not reflect real-life behavior.  
Our work is clearly motivated by the criticisms made in this work.

\begin{table*}[th]
\caption{Characteristics of benchmarking datasets.}
\label{tableMetricsEle}
\centering
\begin{tabular}{|l|l|r|r|r|r|r|r|r|r|}
\hline
 & \multicolumn{1}{c|}{Dataset} 
 & \multicolumn{1}{c|}{$DS$} 
 & \multicolumn{1}{c|}{$AS$} 
 & \multicolumn{1}{c|}{$ATS$} 
 & \multicolumn{1}{c|}{$MTS$}
 & \multicolumn{1}{c|}{$F1$}  
 & \multicolumn{1}{c|}{$GGD$} 
 & \multicolumn{1}{c|}{$H_1$} 
 & \multicolumn{1}{c|}{$H_2$}    \\
 & & & & & & \multicolumn{1}{c|}{(\%)} & \multicolumn{1}{c|}{(\%)} & &\\
\hline
1. & \texttt{forests} & 246 & 206 & 61.26 & 162 & 29.73 & 89.88 & 7.07 & 13.24    \\
2. & \texttt{bogPlants} & 377 & 315 & 14.65 & 39 & 4.65  & 16.57  &  6.56 & 11.56 \\
3. & \texttt{chess} & 3196 & 75  & 37.00 & 37 & 49.33 & 93.05  & 5.81 & 10.57 \\
4. & \texttt{foodmart} & 4141 & 1559 & 4.42 & 14 & 0.28 & 3.18  & 10.55 & 15.21 \\
5. & \texttt{mushroom} & 8124 & 119 & 23.00 & 23 & 19.33 & 50.24  & 5.95 & 10.61 \\
6. & \texttt{pumsb} & 49046 & 2113 & 74.00 & 74 & 3.50 & 23.93 & 7.67 & 14.17 \\
7. & \texttt{pumsbStar} & 49046 & 2088  & 50.48  &  63 & 2.42 & 22.15  & 7.76 & 14.19 \\
8. & \texttt{bmsWebview1} & 59602 &  497 & 2.51 & 267 & 0.51 & 51.90  & 7.85 & 14.84 \\
9. & \texttt{connect} & 67557     & 129  & 43.00  &  43 & 33.33 & 82.69  & 6.12 & 11.18 \\
10. & \texttt{bmsWebview2} &    77512     &  3340 & 4.62   & 161 & 0.14 & 12.95  & 10.46 & 18.07 \\
11. & \texttt{belgiumRetail} &    88162     &  16470 & 10.31 & 76 & 0.06 & 2.65  & 11.27 & 20.45\\
12. & \texttt{skin} & 245057     & 12  & 3.99  &  4 & 33.30 & 84.85  & 3.36 & 5.01  \\
13. & \texttt{accidents} & 340183  & 468  & 33.81 & 51 & 7.22 & 42.84  & 6.49 & 11.91 \\
14. & \texttt{onlineRetail} & 541909     & 2604  & 4.36  &  8 & 0.17 & 0.53  & 8.91 & 12.59 \\
15. & \texttt{recordLink} &    574913 & 27  &  10.00  &  10 & 37.04 & 78.63  & 3.80 & 6.43 \\
16. & \texttt{kosarak} & 990002     &  41270 & 8.09 & 2498 & 0.02 & 3.89  & 10.43 & \\
17. & \texttt{kddcup99} & 1000000  & 135  & 16.00 & 16 & 11.85 & 38.93  & 5.04 & 8.54 \\
18. & \texttt{pamp} & 1000000 & 82  &     23.93 & 26 & 29.19 & 86.57  & 5.48 & 9.85 \\
19. & \texttt{uscensus} & 1000000  & 316  & 48.00 & 48 & 15.19 & 83.92  & 7.14 & 12.99 \\
20. & \texttt{powerc} & 1040000     & 125  & 7.00 & 7 & 5.60 & 48.97  & 4.25 & 7.07 \\
21. & \texttt{chainstore} & 1112949 &  46086 & 7.23  &  170 & 0.02 & 2.84  & 12.96 & \\ \hline 
\end{tabular}
\end{table*}

\paragraph{On database characterization.}

A central property of transactional databases is their density. The best intuitive notion of dataset density is given by \citet{Gouda:2001:EMM:645496.658047} who explain that having long frequent itemsets at high levels of support is what makes a dataset dense. This is why frequent mining algorithms typically adapt their method tactics according to this feature since it is more difficult to work through a dense dataset than a sparse one.

Among the works proposing new metrics for the evaluation of transactional databases properties, \citet{Gouda:2001:EMM:645496.658047} is the first one, to the best of our knowledge, to introduce a classification metric for dataset density which is based mainly on the positive border length distribution. Essentially, the authors study the shape of the positive border distribution cut off at specific levels of support and classify datasets into four distinct types. Even though this work sheds some light on the characterization of datasets, it does not give any notion of how the proposed classification behaves over different levels of support. 

Other works that use positive borders for the characterization of databases are
\citet{Ramesh:2003:FID:773153.773181,Ramesh:2005:DSD:1099545.1100406}. Instead of using them to expand our understanding of dataset properties, they focus on the problem of synthetic database generation. For this, multiple levels of positive borders distributions are extracted from an original dataset and transferred to a synthetic one afterwards.

\citet{Palmerini:2004:SPT:967900.968009} propose new measures to learn a dataset density based on the information theoretic concept of entropy and the average support of frequent itemsets. This work uses entropy to take a glance at the database density without going though the inspection of the entire database which is convenient in helping algorithms to decide the best action to take at runtime.

\citet{flouvat2005thorough,Flouvat2010} introduce a dataset classification which follows the idea of positive borders of \citet{Gouda:2001:EMM:645496.658047} described above. The difference is that \citet{Flouvat2010} classifiy dataset density considering negative border length distribution along with that of positive borders. The authors find that negative borders provide extra information which allows a finer understanding of datasets. Concretely, an algorithm's performance is affected by how separated the mean distance between both borders is.

All of the metrics mentioned in these related works are included in our list of proposed metrics, as well as some others. The next section details the different metrics and related concepts considered in our study.

\section{Definition of metrics}
\label{sectionDefintionMetrics}

We propose to carry out different measurements on the \textit{FIMI} and \textit{SPMF} benchmark datasets utilizing metrics found in the literature and new ones of our own. 

We start by defining basic properties of a transactional database such as \textbf{\textit{dataset size (DS)}} which is simply its number of transactions, \textbf{\textit{average transaction size (ATS)}} as the name suggests is the average of all transaction sizes, and \textbf{\textit{maximum transaction size (MTS)}} is the maximum size value of all transactions.

Most of the following metrics are based on $\mathit{FIM}$. Let $\mathit{I}$ be a finite set of different elements called items and let $\mathit{|I|}$ represent its size. In other words, $\mathit{I}$ represents the database's alphabet and hereafter we call the \textbf{\textit{alphabet size}} as \textbf{\textit{(AS)}}. Any subset of $\mathit{I}$ is denoted as an itemset $\mathit{X}$. In particular, an itemset with size $k$ is regarded as a $k$-itemset. Let $\mathit{D}$ be a transactional database of size $\mathit{|D|}$ in which each transaction is represented by an itemset. Duplication of transactions may exist; thus, transactions are differentiated via identifiers.

The support of an itemset $\mathit{sup(X)}$ is defined as the cardinality of the set of transactions in $\mathit{D}$ in which $\mathit{X}$ is a subset. $\mathit{X}$ is considered frequent if its support is greater than or equal to a minimum support $minsup$ defined by the user, i.e., $\mathit{sup(X)} \geq minsup$. This allows to define $\mathit{FI(minsup)}$ or simply $\mathit{FI}$ as the set of all frequent itemsets with support greater or equal to $\mathit{minsup}$.

\cite{durand:hal-01465110} present an intuitive view of \textit{frequent itemset} borders where the positive border is the set of its maximal frequent itemsets (Equation~\ref{pos_border}) and the negative border is the set of minimal infrequent itemsets (Equation~\ref{neg_border}). 
\begin{align} 
	Bd^+(FI) &=\{X \in FI \ \vert \ \forall Y \supset X, Y \not\in FI\} \label{pos_border}\\
    Bd^-(FI) &=\{X \in 2^I \setminus FI \ \vert \  \forall Y \subset X, Y \in FI\} \label{neg_border}
\end{align}

\begin{table*}[th]
\caption{Characteristics of benchmarking datasets which are calculated using the levels of support $S$ defined for each dataset.}
\label{tableMetricsSop}
\centering
\begin{tabular}{|l|l|r|l|r|r|r|r|r|r|r|}
\hline
 & \multicolumn{1}{c|}{Dataset} 
 & \multicolumn{1}{c|}{$MSS$}  
 & \multicolumn{1}{c|}{Levels of support $S$} 
 & \multicolumn{1}{c|}{$MCD$}  
 & \multicolumn{1}{c|}{$ASD$} 
 & \multicolumn{1}{c|}{$FAL$}  
 & \multicolumn{1}{c|}{$PBC$}  
 & \multicolumn{1}{c|}{$PBL$} \\
& & \multicolumn{1}{c|}{(\%)} & \multicolumn{1}{c|}{(\%)} & & & & & \\
\hline
1. & \texttt{forests} & 93.09 &  $\langle 30, 40, 50,\ldots, 90 \rangle$ & 519.69 & 3805.10  & 242.07 & 544.74 & 270.20  \\
2. & \texttt{bogPlants} & 65.25 &  $\langle 10, 20, 30,\ldots, 60 \rangle$ & 699.62 & 2151.31 & 81.33 & 867.58 & 91.87  \\
3. & \texttt{chess} & 99.97 &  $\langle 20, 30, 40,\ldots, 90 \rangle$ & 656.69 & 4222.48 & 544.58 & 870.26 & 681.04 \\
4. & \texttt{foodmart} & 0.60 &  $\langle 0.1, 0.2, \ldots, 0.6 \rangle$ & 18.36 & 0.21 & 0.50 & 18.36 & 0.50 \\
5. & \texttt{mushroom} & 100.00 &  $\langle 10, 20, 30,\ldots, 90 \rangle$ & 599.48 & 4796.45 & 301.26 & 1022.85 & 396.58 \\
6. & \texttt{pumsb} & 99.79 &  $\langle 50, 60, 70, 80, 90 \rangle$ & 634.85 & 2944.61 & 349.69 & 770.19 & 391.60  \\
7. & \texttt{pumsbStar} & 79.01 &  $\langle 30, 40, 50, 60, 70 \rangle$ & 565.21 & 2178.56 & 188.54 & 861.111 & 192.68  \\
8. & \texttt{bmsWebview1} & 6.14 &  $\langle 1, 2, 3, 4, 5, 6 \rangle$ & 106.49 & 23.70 & 5.11 & 111.94 & 5.12 \\
9. & \texttt{connect} & 99.88 &  $\langle 40, 50, 60,\ldots, 90 \rangle$ & 836.10 & 3510.22 & 488.85 & 1765.89 & 717.68  \\
10. & \texttt{bmsWebview2} & 4.86 &  $\langle 1, 2, 3, 4 \rangle$ & 70.99 & 10.23 & 3.17 & 75.37 & 3.16  \\
11. & \texttt{belgiumRetail} & 57.48 &  $\langle 10, 20, 30, 40, 50 \rangle$ & 1444.44 & 1860.98 & 48.89 & 1400.00 & 64.00 \\
12. & \texttt{skin} & 79.25 &  $\langle 10, 20, 30,\ldots, 70 \rangle$ & 1490.00 & 3348.70 & 81.69 & 2714.29 & 99.29 \\
13. & \texttt{accidents} & 99.99 &  $\langle 10, 20, 30,\ldots, 90 \rangle$ & 601.27 & 4526.47 & 419.98 & 661.9 & 585.52  \\
14. & \texttt{onlineRetail} & 10.04 &  $\langle 1, 2, 3,\ldots, 9 \rangle$ & 126.07 & 56.31 & 8.69 & 155.63 & 8.83 \\
15. & \texttt{recordLink} & 99.99 &  $\langle 10, 20, 30,\ldots, 90 \rangle$ & 1803.25 & 5200.77 & 309.71 & 8583.33 & 508.19 \\
16. & \texttt{kosarak} & 60.75 &  $\langle 10, 20, 30,\ldots, 60 \rangle$ & 1888.89 &  2434.53 & 59.83 & 4250.00 & 72.50  \\
17. & \texttt{kddcup99} & 79.36 &  $\langle 10, 20, 30,\ldots, 70 \rangle$ & 1366.71 & 2893.80 & 373.32 & 4000.00 & 454.31 \\
18. & \texttt{pamp} & 94.51  &  $\langle 10, 20, 30,\ldots, 90 \rangle$ & 658.54 & 4499.74 & 425.66 & 895.82 & 471.18  \\
19. & \texttt{uscensus} & 88.23  &  $\langle 30, 40, 50,\ldots, 80 \rangle$ & 546.55 & 2971.37 & 210.29 & 890.65 & 233.58  \\
20. & \texttt{powerc} & 96.74 &  $\langle 10, 20, 30,\ldots, 90 \rangle$ & 1372.14 & 5406.29 & 168.90 & 2125.00 & 236.17\\
21. & \texttt{chainstore} & 5.73 &  $\langle 1, 2, 3, 4, 5 \rangle$ & 100.00 & 16.64 & 4.00 & 100.00 &  4.00  \\ \hline 
\end{tabular}
\end{table*}

\textbf{\textit{Maximum singleton support (MSS)}} \cite{Ramesh:2005:DSD:1099545.1100406} is the maximum over all singleton supports, i.e., $MSS = \max_{\forall X \in I} sup(X)$. 
This metric provides an upper bound of the support of any itemset and therefore constitutes an important parameter.
This metric also guides us in choosing appropriate thresholds of min support values. That is, $S$ is the sequence of $minsup$ values given by the user in a itemset mining operation. Namely, $S = \langle s_1, s_2, s_3,\dots{}, s_m\rangle$ where $0 < s_k < s_{k+1}$ and $s_m \le MSS$.

\textbf{\textit{Maximum cardinality difference (MCD)}} is defined to approximate how the number of frequent itemsets is distributed throughout the range of possible supports $S$ defined above. That is, we defined \textit{MCD} as the area under the curve defined by the set of ordered pairs $\{(s, \vert \mathit{FI(s)} \vert ) \ \vert \ \forall s \in S \}$. After normalization, a low \textit{MCD} value indicates that the greater percentage of frequent itemsets concentrates at the lowest levels of support whereas a higher value suggests the number of frequent itemsets does not variate abruptly throughout the different support levels $S$. 

Similar to \textit{MCD}, \textbf{\textit{positive border cardinality (PBC)}} is the area under the curve given by the set of ordered pairs $\{(s, \vert \mathit{Bd^+(FI(s))} \vert ) \ \vert \ \forall s \in S \}$. Here, we are interested in knowing how the cardinality of the positive border set (Equation \ref{pos_border}) changes over the different levels of support. 

Likewise, \textbf{\textit{negative border cardinality (NBC)}} is based on the same formulation as $PBC$, but instead it utilizes the negative border set (Equation \ref{neg_border}) to define the set of points $\{(s, \vert \mathit{Bd^-(FI(s))} \vert ) \ \vert \ \forall s \in S \}$.

\textit{\textbf{Gaifman graph density (GGD)}} is another metric we use to approximate the dataset density. Here, an undirected graph $G = (V, E)$ called Gaifman graph is built from a transactional database $\mathit{D}$. Its vertices are items (i.e. $V = I$) and two items share an edge if they appear in at least a transaction together.
%
Then, we compute the density of the Gaifman graph $G$ as the ratio of the cardinality of its set of edges $\vert E \vert$ to the maximum number of edges, that is $GGD = \frac{2 \ \mathit{\vert E \vert}}{\mathit{\vert I \vert} \ (\mathit{\vert I \vert} - 1)}$.

\cite{Palmerini:2004:SPT:967900.968009} propose three metrics.
The first one, \textbf{\textit{fraction of 1s (F1)}}, represents the dataset $D$ as a binary matrix $D'$ where every row is deemed as a transaction of $D$ and every column as an item $X \in I$. An element in $D'$ is marked by 1 depending on whether item $X$ appears in the transaction, and 0 otherwise. The fraction of 1s is calculated as the ratio of the number of 1s in $D'$ to its number of elements, i.e., $F1 = \frac{cardinality\ of\ 1s}{\mathit{\vert D \vert} \ \mathit{\vert I \vert}}$.

The second one, \textit{FI} average support, as its name suggests takes the result of a mining process\textemdash $\mathit{FI(minsup)}$ along with every frequent itemset support\textemdash and measures the distance between the average support of all \textit{FI} itemsets and $minsup$. We take this idea, however, we here consider \textbf{\textit{average support distance (ASD)}} as the area under the curve formed by the points $\{(s, \gamma_{s}) \ \vert \ \forall s \in S\}$ where $\gamma_{s} = \frac{1}{\vert FI(s) \vert} \sum_{\forall X \in FI(s)} sup(X)$. Hence, it is assumed that a database is denser the greater its \textit{ASD} value is.

\begin{figure}[th]
  \centering
  \includegraphics[width=0.41\textwidth]{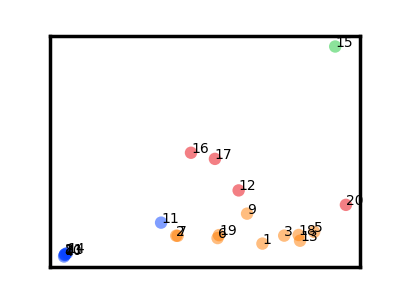}
  \caption{Scatter plot of our 21 datasets using metrics $ASD$ (x axis) and $PBC$ (y axis). The four clusters found by \texttt{K-Means} are color-marked.}
  \label{figure_tsne}
\end{figure}

We define \textbf{\textit{FI average length (FAL)}} using a procedure similar to \textit{ASD}. However, instead of considering the average support of all $FI$ itemsets, here \textit{FAL} computes the area under the curve using the average length of all $FI$ itemsets for every $minsup$ defined by the user. For this, we define the set of points as $\{(s, \lambda_{s}) \ \vert \ \forall s \in S\}$ where $\lambda_{s} = \frac{1}{\vert FI(s) \vert} \sum_{\forall X \in FI(s)} \vert X \vert$. Recall the concept of the sequence of $minsup$ $S$ is explained above.

In addition, \textbf{\textit{positive border average length (PBL)}} and \textbf{\textit{negative border average length (NBL)}} are denoted with the same formulation as the previous \textit{FAL} metric. In this case, \textit{PBL} and \textit{NBL} utilize the set $Bd^+(FI(s))$ and $Bd^-(FI(s))$ respectively instead of the $FI(s)$. It is worth mentioning that in this work all the metrics which base their calculation on itemset lengths such as \textit{FAL}, \textit{PBL}, and \textit{NBL} are not normalized. 


For their third metric, \citet{Palmerini:2004:SPT:967900.968009} use the concept of entropy over probability distributions of $k$-itemsets in order to attempt a density estimation of the database without needing to work through it entirely. This is done by first computing $H_1$ which is the entropy of the $1$-itemset set (i.e. the singleton set), then $H_2$ considering the $2$-itemset set, and so forth. 

In general they define
$H_k = -\sum_{i \in \alpha_k} { p_i \log_2(p_i)},$
where $\alpha_k$ is the set of $k$-itemsets, and $p_i$ corresponds to the relative support of a given $k$-itemset.
Datasets with low entropies are considered denser.

We want to point out that most metrics have been described and used in the past, however, we have defined new ones ($MCD, GGD$) and adapted existing ones to our needs ($ASD, FAL, PBL, PBC, NBC, NBL$).


\begin{figure}[th]  
  \centering
  \includegraphics[width=0.41\textwidth]{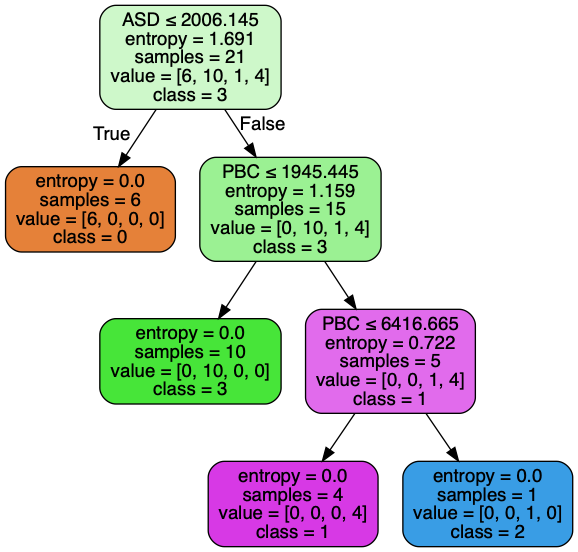}
  \caption{Decision tree classifying datasets into the four clusters found by \texttt{K-Means}.}
\label{fig: decisionTree}
\end{figure}

\section{Dataset characterization}
\label{results}

In this section we present the characterization of the datasets under study, and introduce our notion of \emph{minimum representative benchmark}. Finally, we include a description of how the \textit{FIM} literature has employed these datasets for algorithm benchmarking.

We consider two public repositories: 
\textit{FIMI}\footnote{\url{http://fimi.ua.ac.be} (accessed September 1, 2017)}, and \textit{SPMF}\footnote{\url{http://philippe-fournier-viger.com/spmf/ } (accessed September 1, 2017)}; these two repositories have been made available to the community for the purpose of benchmarking new and existing \textit{FIM} algorithms. Both repositories possess collections of real-life datasets (listed in Table~\ref{tableMetricsEle}, from row 3 to row 21). The first two datasets (i.e. row 1 and 2) are also real datasets taken from W. Hamalainen of the University of Eastern Finland\footnote{\url{http://www.cs.uef.fi/\~whamalai/datasets.html} (accessed September 1, 2017)}. 
The first four columns of Table~\ref{tableMetricsEle} present elemental properties of any transactional database: the number of transactions \textit{(DS)}, alphabet size \textit{(AS)}, average transaction size \textit{(ATS)}, and the maximum transaction size \textit{(MTS)}. 

\begin{table*}[th]
\caption{Benchmarking datasets utilized in published empirical studies. Columns with a citation correspond to the studies, and an ``x'' marks the fact that the dataset has been included in the study. The datasets have been grouped by clusters according to Figure \ref{figure_tsne}.}
\label{tableLiterature}
\begin{center}
\begin{tabular}{|l|l|c|c|c|c|c|c|c|c|c|c|c|c|}
\hline
 & \textsl{Dataset} 
 & \textsl{Cluster}  
& \textsl{\cite{DBLP2003}} 
& \textsl{\cite{Uno2003LCMAE}} 
& \textsl{\cite{Burdick2003MAFIAAP}} 
& \textsl{\cite{Liu03afopt:an}}  
& \textsl{\cite{Goethals03surveyon}} 
& \textsl{\cite{1501819}} 
& \textsl{\cite{Deng2012}} 
& \textsl{\cite{DENG20144505}} 
& \textsl{\cite{Han2004}}
& \textsl{\cite{989550}}
& \textsl{\cite{Uno2004LCMV2}}
\\
\hline
4. & \texttt{foodmart}       & 0 &   &   &   &   &   &   &   &   &   &   & \\
8. & \texttt{bmsWebview1}    & 0 & x & x & x & x & x &   &   &   &   & x & x\\
10. & \texttt{bmsWebview2}   & 0 & x & x & x &   &   &   &   &   &   &   & x\\
11. & \texttt{belgiumRetail} & 0 & x &   &   &   &   &   & x &   &   &   & x\\
14. & \texttt{onlineRetail}  & 0 &   &   &   &   &   &   &   &   &   &   & \\
21. & \texttt{chainstore}    & 0 &   &   &   &   &   &   &   &   &   &   & \\ 
\hline
12. & \texttt{skin}          & 1 &   &   &   &   &   &   &   &   &   &   & \\
16. & \texttt{kosarak}       & 1 & x &   &   &   &   & x &   &   &   &   & x\\
17. & \texttt{kddcup99}      & 1 &   &   &   &   &   &   &   &   &   &   & \\
20. & \texttt{powerc}        & 1 &   &   &   &   &   &   &   &   &   &   & \\
\hline
15. & \texttt{recordLink}    & 2 &   &   &   &   &   &   &   &   &   &   & \\
\hline
1. & \texttt{forests}        & 3 &   &   &   &   &   &   &   &   &   &   &\\
2. & \texttt{bogPlants}      & 3 &   &   &   &   &   &   &   &   &   &   & \\
3. & \texttt{chess}          & 3 & x & x & x & x &   &   &   &   &   &   & x\\
5. & \texttt{mushroom}       & 3 & x & x & x & x & x &   &   & x &   &   & x\\
6. & \texttt{pumsb}          & 3 & x & x & x & x &   &   & x &   &   &   & x\\
7. & \texttt{pumsbStar}      & 3 & x & x & x &   &   & x &   &   &   &   & x\\
9. & \texttt{connect}        & 3 & x & x & x & x &   & x &   & x & x &   & x\\
13. & \texttt{accidents}     & 3 & x &   &   &   &   & x & x &   &   &   & x\\
18. & \texttt{pamp}          & 3 &   &   &   &   &   &   &   &   &   &   & \\
19. & \texttt{uscensus}      & 3 &   &   &   &   &   &   &   &   &   &   & \\
\hline
\end{tabular}
\end{center}
\end{table*}

Metrics \textit{F1} and \textit{GGD} of Table~\ref{tableMetricsEle} and \textit{MSS} of Table~\ref{tableMetricsSop} are presented as percentage. Metrics \textit{H1} and \textit{H2} of Table~\ref{tableMetricsEle} have been calculated based on the formulation of entropy \textit{H} using $1$-itemset and $2$-itemset sets, respectively. The five metrics of Table~\ref{tableMetricsSop}\textemdash \textit{MCD, ASD, FAL, PBC,} and, \textit{PBL}\textemdash are computed at different levels of support. These levels of support are presented in the same table for each dataset  and are named by the notation $S$ defined previously in Section~\ref{sectionDefintionMetrics}. 

The formulation of the metrics related to negative borders \textit{NBC} and \textit{NBL} have been presented in this work, however, the experimental results on these metrics will be included in a future work. 

Two observations are important. Firstly, each of the support levels in $S$ is given in percentage and is denoted as the ratio of the number of transactions to the size of the database. Secondly, that the highest level of support used with a particular dataset is bounded by the datasets' \textit{MSS} value. Given this upper bound, we have taken (roughly) equidistant intervals of support.

Next, we show in Table~\ref{tableMetricsEle} and Table~\ref{tableMetricsSop} the values obtained for each elemental metric and also for the more sophisticated metrics described in Section~\ref{sectionDefintionMetrics} for our 21 datasets.

\subsection{Minimum representative benchmark.}
\label{subsec:mrb}

Intuitively, a representative benchmark is a set of datasets one can safely do empirical studies on (for example, checking whether one algorithm is better than another in a benchmarking study). As an example, suppose that we only care about the number of transactions of a dataset. Then, a benchmark would be representative if it were to include datasets with few transactions as well as datasets with a high number of them, and perhaps datasets with an intermediate number of transactions. In a sense, we are clustering the datasets according to this particular value, and a representative benchmark is one that includes at least a representative from each cluster.
We follow this idea, but instead of using a single characteristic, we use a whole array of them -- namely, all the metrics considered in Section~\ref{sectionDefintionMetrics}.

Therefore, our working assumption is that if the values of the benchmark cover the whole range of possible values, then it is representative. Moreover, we seek \emph{minimum} benchmarks in the sense that we want to include as few datasets as possible (adding datasets with similar characteristics to the benchmark does not enrich the benchmark and slows down the benchmarking process).

We have clustered the 21 datasets used in this paper into four clusters using \texttt{K-Means}.
In order to perform the clustering, we have used the \texttt{scikit-learn} package \cite{scikit-learn}, more concretely its \texttt{kmeans++} initialization version with 500 restarts. 
Metrics have been scaled prior to clustering with \texttt{RobustScaler} from \texttt{scikit-learn} to avoid  different ranges of values perverting the clustering process.
Finally, we have chosen to partition into four clusters because this resulted into the most succint description of the resulting clusters (please see paragraph below on the tree description of the clusters regarding Figure~\ref{fig: decisionTree}).

Figure~\ref{figure_tsne} shows the four clusters found, which are: $\{4, 8, 10, 11, 14, 21\}$ (cluster 0), $\{12, 16, 17, 20\}$ (cluster 1), $\{15\}$ (cluster 2), and finally $\{1, 2, 3, 5, 6, 7, 9, 13, 18, 19\}$ (cluster 3). The underlying idea is that datasets within clusters are similar to each other (in terms of their metrics' values), but different to others in different clusters. Any representative benchmark should therefore include at least one dataset of each of the four clusters found (a representative benchmark constitutes a hitting set for the four clusters).

In order to understand the nature of the four clusters found by \texttt{K-Means}, we have generated a decision tree that is able to classify all datasets into their right cluster using metrics' values as attributes. This tree can be seen in Figure~\ref{fig: decisionTree}. Cluster 0, for example, is formed by datasets having a value of at most 2006 in the $ASD$ metric. Cluster 1 consists of those datasets having large value in $ASD$ and a value for $PBC$ between 1945 and 6416. Cluster 3 consists of datasets having large values in $ASD$ but small values for $PBC$. Finally, cluster 2 consists of datasets having large values for $ASD$ and $PBC$.

\subsection{Literature review of empirical studies.}


Table \ref{tableLiterature} presents the real-life benchmarking datasets used by several authors to performing comparison studies on $\mathit{FIM}$ algorithms. In here, \citet{DBLP2003} and \citet{Uno2004LCMV2} based their benchmarking studies on most of the datasets presented in this work and closely followed by \citet{Uno2003LCMAE} and \citet{Burdick2003MAFIAAP}. This information allows us to identify which works need to expand their pool of benchmarking datasets in order to consider their outcome as closer to reality, in addition to giving the $\mathit{FIM}$ community a global vision of the way the set of public benchmarking datasets is distributed among the different authors.

In order to find out which of the sets employed in the empirical studies of Table~\ref{tableLiterature} are \emph{representative}, we need to establish which of the benchmarks employed constitute hitting sets of the four groups defined previously in Section~\ref{subsec:mrb}.

The datasets used in \citet{DBLP2003} and \citet{Uno2004LCMV2} are $\{3, 5, 6, 7,  8, 9, \allowbreak 10,  11, 13,  16\}$. This set intersects with all the clusters with the exception of \texttt{recordLink} dataset (cluster 2) thus they miss the characteristic metrics provided by this cluster. So, both studies would only need to include \texttt{recordLink} to comply with the requirement. The remaining studies mentioned in Table~\ref{tableLiterature} miss datasets from more than one cluster. 

Hence, according to the analysis and criteria used in this work, we conclude that there is no study that uses a representative benchmark which is very useful for the various authors to take into account when deciding the set of benchmarking datasets to be used in their experiments. Besides, this work serves as a guide and motivation for new benchmarking datasets to be introduced into the public domain in order to be further characterized and extend the range of characteristics coverage.


\section{Conclusions and future work}
\label{conclusion}

In this work we provide a study of metrics for the characterization of transactional databases commonly used for the benchmarking of itemset mining algorithms. We argue that these metrics are needed in order to assess the diversity and, therefore, the representativeness of such benchmarks so that conclusions drawn from benchmarking studies are sufficiently supported. We study the representativeness of databases used as the basis for existing benchmarking studies found in the literature based on our characteristics, thus assessing which of the studies has a better support for their claims. Finally, we propose a way of obtaining benchmarks with guaranteed diversity that authors of new benchmarking studies can benefit from.

As future lines of research we are always on the lookout for new metrics used for the characterization of databases. As new metrics and databases are incorporated into our lists, we should revise the \emph{minimum representative benchmarks} accordingly. This could be thought of as an evolving process and as such could (and should) potentially be included in the $\mathit{FIMI}$ or similar repository to be used by the data mining community.

Finally, a line of research that we are pursuing is that of synthetic database generation. The existence of this paper is, in fact, due to the fact that we identified the need for characterizing databases so that we could assess in some ways the datasets that we generate. With characterization metrics, we can check the nature of the databases that we generate (generally mimicking some original database that due to privacy and ownership constraints cannot be shown or used). For example, we could check whether synthetic and original are similar enough, if that is what is desired. Or we could study how parameter tuning in the generative algorithms affects the nature of the generated databases. In any case, being able to characterize databases is a powerful and necessary tool to understand the nature of the generated datasets.

\bibliographystyle{plainnat}
\bibliography{bibliography}
\end{document}